\documentclass[preprint,12pt]{elsarticle}

\usepackage{amssymb}
\usepackage{amsmath}

\newcounter{bla}

\journal{Computer Physics Communications}

\usepackage{url}
\usepackage{multirow}
\usepackage{graphicx}
\usepackage{color}
\usepackage{tikz}
\usepackage{pgfplots}
\pgfplotsset{compat=1.9}

\usepackage{xspace}

\usepackage{url}
\usepackage{subcaption}

\begin{document}

\begin{frontmatter}
\title{DASHMM Accelerated Adaptive Fast Multipole Poisson-Boltzmann Solver
on Distributed Memory Architecture}

\author[a]{B. Zhang\corref{author}}
\author[a]{J. DeBuhr}
\author[b]{D. Niedzielski}
\author[c]{S. Mayolo}
\author[d]{B. Lu}
\author[a]{T. Sterling}

\cortext[author]{
Corresponding author.\\
\textit{E-mail address:} zhang416@indiana.edu
}
\address[a]{
  Center for Research in Extreme Scale Technologies, School of Informatics
  Computing, and Engineering, Indiana University, Bloomington, IN, 47404, USA}
\address[b]{
  Department of Physics,
  Rensselaer Polytechnic Institute, Troy, NY, 12180, USA
}
\address[c]{
  Department of Computer Science,
  Tennessee Technological University, Cookeville, TN, 38505, USA
}
\address[d]{
  State Key Laboratory of Scientific/Engineering Computing, Academy of
  Mathematics and Systems Science, Chinese Academy of Sciences, Beijing, 100190,
  China
}

\begin{abstract}
We present an updated version of the AFMPB package for fast calculation of
molecular solvation-free energy. The main feature of the new version is the
successful adoption of the DASHMM library, which enables AFMPB to operate on
distributed memory computers. As a result, the new version can easily handle
larger molecules or situations with higher accuracy requirements. To
demonstrate the updated code, we applied the new version to a dengue virus
system with more than one million atoms and a mesh with approximately 20
million triangles, and were able to reduce the time-to-solution from 10 hours
reported in the previous release on a shared memory computer to less than 30
seconds on a Cray XC30 cluster using $12,288$ cores.
\end{abstract}


\end{frontmatter}

{\bf PROGRAM SUMMARY/NEW VERSION PROGRAM SUMMARY}

\begin{small}
\noindent
{\em Program Title:} Distributed AFMPB \\
{\em Licensing provisions:} GNU General Public License, version 3\\
{\em Programming language:} C++ \\
{\em Supplementary material:} \\
{\em Journal reference of previous version:} AEGB\_v2\_0\\
{\em Does the new version supersede the previous version?:} Yes\\
{\em Reasons for the new version:} Memory requirement and time-to-solution for
large molecules of interest exceed the capacity of shared memory computers. \\
{\em Summary of revisions:} The revision extends the AFMPB to operate on
distributed memory computers using the DASHMM [1, 2] library. \\
{\em Nature of problem:} Numerical solution of the linearized Poisson-Boltzmann
equation that describes electrostatic interactions of molecular systems in 
ionic solutions. \\
{\em Solution method:} The linearized Poisson-Boltzmann equation is reformulated 
as a boundary integral equation and is subsequently discretized using the
node-patch scheme. The resulting linear system is solved using GMRES. 
 Within each iteration, the matrix-vector multiplication is
accelerated using the DASHMM library on distributed memory computers. 


\end{small}

\section{Introduction}
\label{sec:intro}

The Poisson-Boltzmann (PB) continuum electrostatic model has been adopted
in many simulation tools for theoretical studies of electrostatic
interactions between biomolecules such as proteins and DNAs in aqueous
solutions. Various numerical techniques have been developed to solve the
PB equations and help elucidate the electrostatic role in many
biological processes, such as enzymatic catalysis, molecular recognition
and bioregulation. Packages such as DelPhi~\cite{delphi,delphi2},
MEAD~\cite{Bashford1997}, UHBD~\cite{uhbd, Madura1995},
PBEQ~\cite{PBEQ}, ZAP~\cite{Grant2001}, and MIBPB~\cite{mibpb}
are based on finite-difference methods. Packages such as the Adaptive
Poisson-Boltzmann Solver (APBS)~\cite{apbs} is based on finite volume/multigrid
methods. In a circumstance where the linearized PB is applicable, the partial
differential equations can be reformulated into a set of surface
integral equations (IEs) by using Green's theorem and potential
theory. The unknowns in the IEs are located on the molecular surface,
and the resulting discretized linear system can be solved
very efficiently and accurately with certain fast convolution
algorithms, such as the fast multipole method (FMM). This strategy
has been implemented in the Adaptive Fast Multipole Poisson-Boltzmann~(AFMPB)
solver.

The AFMPB solver was first released as a sequential package written in
Fortran~\cite{Lu2010}. A major update was released in 2015~\cite{afmpb}
that used the Cilk runtime for parallelization on shared memory computers and
provided built-in surface mesh generation capability. Over time, the target
applications of AFMPB grew larger and required higher accuracy, which meant
that it became difficult or impossible to find a shared memory computer with
sufficient memory or processing capacity to solve the problems of interest.
This paper presents a major update to AFMPB v. 2.0 \cite{afmpb} that extends
its operation to distributed memory architectures.

To easily create a distributed version of AFMPB, the Dynamic Adaptive System
for Hierarchical Multipole Method (DASHMM) \cite{debuhr2016, debuhr2017b}
library was adopted as the central driver of the multipole methods used by
AFMPB. DASHMM is built on top of the asynchronous many-tasking HPX-5
\cite{kulkarni16, kissel16} runtime system. It is fine-grained, data-driven,
and has very good strong-scaling. More importantly, the DASHMM programmer
interface is independent of HPX-5 and no knowledge of the runtime is required
from its end-users. As a result, upgrading AFMPB to use DASHMM was simple and
fast.

In addition to the distribution of the FMM computation, the use of GMRES by
AFMPB needed to be upgraded to work in distributed memory architectures.
Before this update, AFMPB used SPARSKIT's~\cite{sparskit} GMRES implementation.
This sequential, Fortran, implementation was replaced with a straightforward,
fully-distributed GMRES patterned after the original SPARSKIT implementation.

These changes allow the new version of AFMPB to handle larger molecules
or situations with higher accuracy requirements that were unfeasible on
shared memory computers.

The organization of this paper is as follows. Section~\ref{sec:bg} reviews
the mathematical models and discretization methods adopted in AFMPB and
where DASHMM can be applied. Section~\ref{sec:dashmm} describes how DASHMM
is integrated into AFMPB. Section~\ref{sec:install} provides installation 
guide and job examples. Section~\ref{sec:result} shows numerical results 
on two molecules. Section~\ref{sec:conclusion} concludes the paper.

\section{Overview of the AFMPB Solver}
\label{sec:bg}

The electrostatic force is considered to play an important role in
the interactions and dynamics of molecular systems in aqueous
solution. In the Poisson equation model, when the charge density that
describes the electrostatic effects on the solvent outside the
molecules is approximated by a Boltzmann distribution, the continuum
nonlinear Poisson-Boltzmann (PB) equation assumes the following form 
\begin{equation}
\label{eq:pb}
-\nabla \cdot (\epsilon \nabla \phi) + \bar{\kappa}^2 \sinh(\phi) = 
\sum_{k=1}^M q_k \delta(r - r_k)
\end{equation} 
In the formula, the molecule is represented by $M$ point charges
$\{q_k\}$ located at $\{r_k\}$, $\epsilon$ is the position-dependent
dielectric constant, $\phi$ is the electrostatic potential at location
$r$, $\bar{\kappa}$ is the modified Debye-H\"{u}ckel parameter, where
$\bar{\kappa} = 0$ in the molecule region and
$\bar{\kappa}=\sqrt{\epsilon}\kappa$ in the solution region, and
$\kappa$ is the inverse of the Debye-H\"{u}ckel screening length
determined by the ionic strength of the solution. When the
electrostatic potentials are small, the linearized Poisson-Boltzmann
(LPB) equation 
\begin{equation}
\label{eq:lpb}
-\nabla \cdot(\epsilon \nabla \phi) + \bar{\kappa}^2 \phi = 
\sum_{k=1}^M q_k \delta (r - r_k)
\end{equation} 
becomes valid, equipped with the interface conditions $[\phi]=0$ (by
the continuity of the potential) and 
$[\epsilon \frac{\partial \phi}{\partial n}]=0$ (by the conservation
of flux). Here, $[\,\,]$ denotes the jump across the molecular surface
and $\frac{\partial}{\partial n}$ is the outward (towards the solvent)
normal direction at the surface. 

Instead of solving (\ref{eq:lpb}) directly, the AFMPB package adopts
an alternative reformulation that maps (\ref{eq:lpb}) into 
\begin{align}
\left (\frac{1}{2\bar{\epsilon}} + \frac{1}{2} \right) f_p 
& = \oint_S \left [ (G_{pt} - u_{pt}) h_t - 
  \left ( \frac{1}{\bar{\epsilon}} \frac{\partial G_{pt}}{\partial n}
  - \frac{\partial u_{pt}}{\partial n} \right) f_t \right ] dS_t \nonumber \\ 
& \qquad + \frac{1}{\epsilon_{ext}} \sum_{k=1}^M q_k G_{pk} \nonumber  \\
\left (\frac{1}{2\bar{\epsilon}} + \frac{1}{2} \right) h_p 
& =  \oint_S \left [ \left ( \frac{\partial G_{pt}}{\partial n_0} - 
  \frac{1}{\bar{\epsilon}} \frac{\partial u_{pt}}{\partial n_0} \right) h_t
  - \frac{1}{\bar{\epsilon}} \left (
  \frac{\partial^2 G_{pt}}{\partial n_0 \partial n} - 
  \frac{\partial^2 u_{pt}}{\partial n_0 \partial n} \right) f_t 
  \right] dS_t + \nonumber \\
& \qquad \frac{1}{\epsilon_{ext}} 
\sum_{k=1}^M q_k \frac{\partial G_{pk}}{\partial n_0} \label{eq:2nd}
\end{align}
where 
\[
G_{pt} = \frac{1}{4\pi |r_t - r_p|}, \quad 
u_{pt} = \frac{e^{-\kappa|r_t - r_p|}}{4\pi |r_t - r_p|}, \quad 
f = \phi^{ext}, \quad h = \frac{\partial \phi^{ext}}{\partial n}, \quad 
\bar{\epsilon} = \frac{\epsilon_{ext}}{\epsilon_{int}}, 
\]
and $n_0$ is the outward unit normal vector at point $p \in S$. This alternative
reformulation is a Fredholm integral equation of the second kind,
which provides the analytic foundation for a well-condition system
of equations with an adequate discretization method. 

\begin{figure}
\centering
\begin{tikzpicture}[yscale=0.45, xscale=0.75]
\draw [fill=blue!10] 
(1,0.5) to (1, 4/3) to (3/2,2) to (7/3,8/3) to (6/2,5/2) 
to (11/3,4/3) to (7/2, 0) to (10/3, -1) to (5/2,-1) to (2,-5/3) to (3/2, -1) 
to (1, -2/3) to (1,0.5); 

\draw [fill=red!10] 
(7/2, 0) to (11/3, 4/3) to (9/2, 3/2) to (16/3, 4/3) to (6, -1/2) to 
(6, -5/3) to (11/2,-5/2) to (14/3, -8/3) to (4, -2) to (10/3, -1) to (7/2,0);

\draw (0,0) to (1,3) to (4,4) to (5,-1) to (3,-3) to (1,-3) to (0,0); 
\foreach \x/\y in {0/0, 1/3, 4/4,5/-1, 3/-3,1/-3} {
  \draw (2,1) to (\x, \y);
}
\node at (2.35, 1) [scale=0.7] {$v_1$};

\draw (4,4) to (7,0) to (6, -4) to (3,-3);
\draw (5, -1) to (7,0);
\draw (5, -1) to (6, -4);
\node at (5, 0) [scale=0.7] {$v_2$}; 

\foreach \y in {8, 6, 4, 2, 0, -2} {
  \draw[dotted] (8, \y) to (10, \y); 
}

\begin{scope}[xshift=350, yshift=150, rotate=75, scale=0.5] 
\draw [fill=green!10] 
(7/2, 0) to (11/3, 4/3) to (9/2, 3/2) to (16/3, 4/3) to (6, -1/2) to 
(6, -5/3) to (11/2,-5/2) to (14/3, -8/3) to (4, -2) to (10/3, -1) to (7/2,0);

\draw (0,0) to (1,3) to (4,4) to (5,-1) to (3,-3) to (1,-3) to (0,0); 
\foreach \x/\y in {0/0, 1/3, 4/4,5/-1, 3/-3,1/-3} {
  \draw (2,1) to (\x, \y);
}

\draw (4,4) to (7,0) to (6, -4) to (3,-3);
\draw (5, -1) to (7,0);
\draw (5, -1) to (6, -4);
\node at (5, 0) [scale=0.5] {$v_3$}; 
\end{scope}
\end{tikzpicture}
\caption{Illustration of the node-patch scheme used to discretize boundary
integral equation (\ref{eq:2nd}) in the AFMPB solver. For each node $v$, 
its patch is enclosed by the edges connecting the centroid of 
the elements which $v$ belongs to and the midpoints of the edges incident at
$v$. When the nodes are far apart, such as $v_1$ and $v_3$, the integrands of 
the integrations (\ref{eq:abcd}) are taken as constant on the patch. 
When the nodes are close, such as $v_1$ and $v_2$, the integrations 
(\ref{eq:abcd}) are computed directly using detailed information of the 
patches.}
\label{fig:node-patch}
\end{figure}
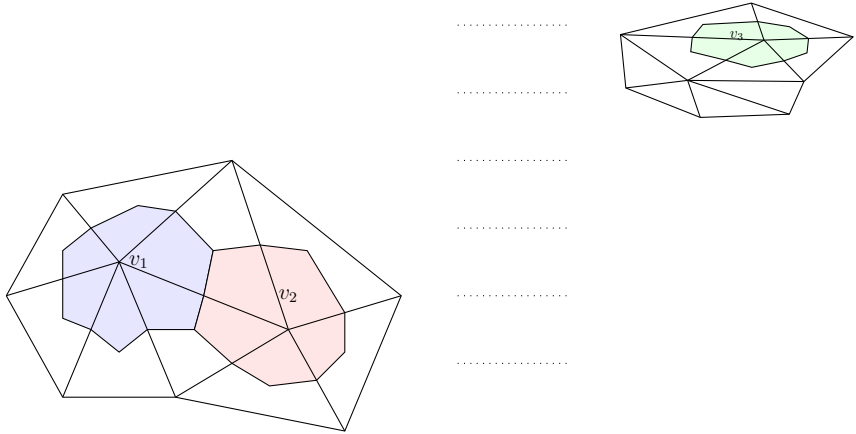

The AFMPB solver adopts the node-patch scheme~\cite{nodepatch} to discretize
the integral equation (\ref{eq:2nd}). The scheme first constructs a patch 
around each node and then assumes unknowns are constant on each patch. 
Specifically, the patch around node $v$, denoted by $\Delta S_v$, 
 is constructed by connecting
the centroid points of the elements that node $v$ belongs to and the
midpoints of the edges that are incident to
$v$. Figure~\ref{fig:node-patch} provides an illustration of the
node-patch approach on mesh of triangular elements. In the figure, the
node-patch for $v_1$ is formed from six surrounding
triangular elements. Using the node-patch scheme, equation (\ref{eq:2nd}) 
can be discretized as 
\begin{align}
\left(\frac{1}{2\bar{\epsilon}} + \frac{1}{2} \right) f_p & = 
\sum_t (A_{pt} h_t - B_{pt} f_t) + \frac{1}{\epsilon_{ext}} \sum_k q_k G_{pk} 
\nonumber \\
\left(\frac{1}{2\bar{\epsilon}} + \frac{1}{2} \right) h_p &= 
\sum_t (C_{pt} h_t - D_{pt} f_t) + \frac{1}{\epsilon_{ext}} \sum_k q_k G_{pk},
\label{eq:disc}
\end{align}
or in matrix-form, 
\begin{equation}
\label{eq:mat}
\begin{bmatrix} 
(\frac{1}{2\bar{\epsilon}} + \frac{1}{2})I + B & -A  \\
D & (\frac{1}{2\bar{\epsilon}} + \frac{1}{2}) I - C 
\end{bmatrix} 
\begin{bmatrix} 
f \\ h
\end{bmatrix} 
= \frac{1}{\epsilon_{ext}}\begin{bmatrix} 
\sum_k q_k G_{pk} \\
\sum_k q_k \frac{\partial G_{pk}}{\partial n_0}
\end{bmatrix} 
\end{equation}
where 
\begin{align}
A_{pt} &= \oint_{\Delta S_t} (G_{pt} - u_{pt}) dS, \nonumber \\
B_{pt} &= \oint_{\Delta S_t}   \left ( 
\frac{1}{\bar{\epsilon}} \frac{\partial G_{pt}}{\partial n}
- \frac{\partial u_{pt}}{\partial n} \right) dS \nonumber \\
C_{pt} &= \oint_{\Delta S_t} \left ( \frac{\partial G_{pt}}{\partial n_0} - 
  \frac{1}{\bar{\epsilon}} 
  \frac{\partial u_{pt}}{\partial n_0} \right) dS \nonumber \\
D_{pt} & = \oint_{\Delta S_t} \frac{1}{\bar{\epsilon}} \left (
  \frac{\partial^2 G_{pt}}{\partial n_0 \partial n} - 
  \frac{\partial^2 u_{pt}}{\partial n_0 \partial n} \right) dS. \label{eq:abcd}
\end{align}
and $I$ is the identity matrix. The discretized system is well-conditioned 
and can be solved efficiently using Krylov subspace methods. For 
(\ref{eq:abcd}), when $p$ and $t$ are far-away, such as $v_1$ and $v_3$ in 
Figure~\ref{fig:node-patch}, the integrands in (\ref{eq:abcd}) are taken
as constants;  When $p$ and $t$ are close, such as $v_1$ and $v_2$ in 
Figure~\ref{fig:node-patch}, each integration is computed directly and 
the computation requires detailed information on the constitution of the patch. 
For this reason, in the existing AFMPB solver, the near-field integrations
are computed only once and saved. 

The overall computation flow of the existing AFMPB solver can be summarized
as follows: 

\begin{enumerate} 
\item Compute the right-hand side of (\ref{eq:mat}) using the Fast Multipole
Method (FMM) with single-layer and double-layer Laplace potentials.
\item For each pair of nodes $p$ and $t$ whose patch integrations 
(\ref{eq:abcd}) are near-field, compute and save $A_{pt}$, $B_{pt}$, $C_{pt}$, 
and $D_{pt}$. 
\item Solve equation (\ref{eq:mat}) using Krylov subspace method. For each 
  matrix-vector multiplication, 
  \begin{enumerate} 
  \item Compute the near-field contribution using the $\{A, B, C, D\}$ 
coefficients in the previous step.  
  \item Compute the far-field contribution with four FMM calls, using 
single-/double-layer Laplace and Yukawa potentials. 
  \end{enumerate}
\end{enumerate} 

\section{DASHMM and AFMPB Integration}
\label{sec:dashmm}

The main feature of the new version is the adoption of the Dynamic
Adaptive System for Hierarchical Multipole Method (DASHMM) library that
enables the AFMPB solver to operate on distributed memory computers. With
the added resources, larger molecules or situations with higher
accuracy requirements that were infeasible on shared memory computers before
can be handled with ease, and the time-to-solution is significantly reduced.
In this section, we first briefly review the features and capabilities
of the existing DASHMM library (version 1.2.0) and then documents the
extensions made to DASHMM during the integration process.

\subsection{DASHMM 1.2.0}

DASHMM is an open-source scientific software library that aims to provide
an easy-to-use system that can provide scalable, efficient, and unified
evaluations of general hierarchical multipole methods on both shared
and distributed memory architectures. The latest public release of DASHMM
is version 1.2.0~\cite{debuhr2017b}. DASHMM leverages the asynchronous
multi-tasking HPX-5 \cite{kulkarni16, kissel16} runtime system for
asynchrony management. HPX-5 defines a broad API that covers most aspects
of the system. Programs are organized as diffusive, message driven
computation, consisting of a large number of lightweight threads and
active messages, executed within the context of a global address space,
and synchronized through the use of lightweight synchronization objects.
The HPX-5 runtime is responsible for managing global allocation,
address resolution, data and control dependence, scheduling lightweight threads
and managing network traffic. Readers interested in more HPX-5 and DASHMM
implementation details are referred to \cite{debuhr2016, debuhr2017b}.
For general end-science users, the DASHMM APIs are completely independent
of HPX-5 and no knowledge of the runtime system is required.
A basic use can be achieved simply through the {\tt Evaluator} object
\begin{verbatim}
dashmm::Evaluator<SourceData, TargetData, Expansion, Method>
\end{verbatim}
that takes source data, target data, expansion, and method as template
parameters.

DASHMM comes with a set of widely used interaction types (kernels), including
the Laplace, Yukawa, and the low-frequency Helmholtz kernels.
It also provides three built-in methods: Barnes-Hut, the classical FMM,
and an variant of FMM that uses exponential expansion~\cite{Greengard1997},
which is called {\tt FMM97} method in the library. The Laplace and Yukawa
kernels and the {\tt FMM97} method are used by the AFMPB solver.

DASHMM also makes a distinction between the mathematical concept of an
expansion, and the concept used in DASHMM. The mathematical concept is a
truncated series of some form (e.g., spherical harmonics) that represents
some potential and is referred as a {\tt View} object.
The concept of an expansion in DASHMM is wider: each expansion in DASHMM
can contain multiple mathematical expansions. In other words, each expansion
in DASHMM is a collection of {\tt View}s, or referred as a {\tt ViewSet}
object. The views of a {\tt ViewSet} can represent the same potential, each
from a different perspective, which is the case for exponential expansions in
the {\tt FMM97} method. The views of a {\tt ViewSet} can also represent
different potentials, which is quite useful in the AFMPB solver. This
means that the four FMM kernel evaluations within each iteration step can be
processed concurrently.

\subsection{New Kernels and Methods}

To complete its computation, AFMPB required the implementation of two kernels
and one method for use with the DASHMM framework. These new kernels and methods
are an example of the extensibility of DASHMM. As long as these new classes
conform to the required interface needed by DASHMM, the library can provide a
parallel distributed FMM evaluation. These new kernels and method are not
included in DASHMM, but rather are application specific extensions usable by
DASHMM.

\textbf{New kernels:} AFMPB solves (\ref{eq:mat}) using a Krylov subspace
method. Applying the left-hand side matrix to each input vector involves
four different kernels. In addition to the single-layer Laplace and Yukawa
kernels already built into the DASHMM library, AFMPB implemented operators
for the double-layer Laplace and double-layer Yukawa kernels. Notice that the
multipole/local expansion for the double-layer potentials share the same form
as the single-layer potentials. This means, only three operators are needed
for each new kernel: (1) {\tt S\_to\_M} operator that generates a multipole
expansion from a set of source points; (2) {\tt S\_to\_L} operator that
generates a local expansion from a set of source points; (3) {\tt L\_to\_T}
operator that evaluates the local expansion at target points. For the reason
discussed next, the {\tt M\_to\_T} operator is not needed.

\textbf{New Method:} In adaptive FMM, each target box $B_t$ is associated
with four types of lists~\cite{Carrier1988} of source boxes. Particularly,
a box $B_s$ is said to be in list-3 of $B_t$ if $B_t$ is well separated
from $B_s$ but is not well separated from the parent of $B_s$. This is very
cost effective because the multipole expansion is used at the earliest
possible time. In the AFMPB solver, however, list-3 boxes have to be processed
the same way as list-1 boxes using the {\tt S\_to\_T} operator to achieve the
required accuracy. This only calls a slight modification of the built-in
{\tt FMM97} method, but it has an impact on the strong-scaling performance
(see Section~\ref{sec:result}).

\subsection{DASHMM extensions}
\label{subsec:ext}

In addition to new kernels and methods to be used by DASHMM, AFMPB drove some
improvements to the DASHMM library. These are outlined below.

\textbf{New API:} In DASHMM 1.2.0, the multipole evaluation is done in
a monolithic way through the single {\tt evaluate} method of the
{\tt Evaluator} class. It first constructs the auxiliary structures, such
as the dual tree and the directed acyclic graph (DAG), and then evaluates
the DAG. This approach provides an easy-to-use and complete evaluation
of a given multipole method. However, this one-size-fits-all approach is
not well tuned for iterative methods that uses the same DAG multiple
times with different input data. To address this, DASHMM now supports
evaluation split into phases, including:
\begin{description}
\item[{\tt create\_tree}] Partition the source and target points into two
trees. The trees can be identical, partially overlapping, or completely
disjoint. In AFMPB, the right-hand side of (\ref{eq:mat}) is a case where
the sources (atoms inside the molecule) are completely disjoint from the
targets (mesh nodes on the molecule surface), and the left-hand side of
(\ref{eq:mat}) is a case where the sources and targets are the same (mesh
nodes on the molecular surface).
\item[{\tt create\_DAG}] This method takes the handle of the tree constructed
from {\tt create\_tree} and a given multipole {\tt method} object to
connect the trees into a DAG.
\item[{\tt execute\_DAG}] This method performs the evaluation of the
multipole method.
\item[{\tt reset\_DAG}] This method resets various DASHMM internal control
objects. Once complete, the DAG is ready to execute a new round of
execution.
\item[{\tt destroy\_DAG}] This method destroys the DAG.
\item[{\tt destroy\_tree}] This method destroys the dual trees.
\end{description}
\noindent By separating the evaluation into phases AFMPB can build the tree and
DAG only once, and evaluate that DAG repeatedly. This saves the overhead of
building an identical tree and DAG for each iteration.

\textbf{New Functionality:} The previous AFMPB solver first generates
the $\{A, B, C, D\}$ coefficients for the near-field integration before
starting the iterative solution phase. This synchronization barrier blocks the
far-field evaluation of the first matrix-vector multiplication of the
iterative solve, and increases the overall execution time. The updated version
of AFMPB removes this barrier. AFMPB directly enters
the iterative solve phase, generates and saves the $\{A, B, C, D\}$
coefficients for future iterations while making progress on the
far-field evaluation.
This suggested additional functionality from DASHMM:
the detailed patch information associated with each node is needed
only in the first iteration and one should avoid communicating unnecessary
large messages. DASHMM addresses this by introducing the {\tt Serializer}
object, which has three member functions:
\begin{description}
\item[{\tt size}] This takes a handle to the object in question and returns
the size in bytes of the serialized object.
\item[{\tt serialize}] This serializes the given object into the buffer
provided and then returns the address after the serialized data.
\item[{\tt deserialize}] This deserializes the given object into the
buffer provided and then returns the address after the data used in the
deserialization.
\end{description}
DASHMM now defines a method, {\tt set\_manager}, on its {\tt Array} type
that allows the array to update its binding to a {\tt Serializer} during the
course of execution. Two {\tt Serializer} objects
are defined in the new version of AFMPB: one serializes detailed patch
information and is used only in the first iteration, and the other
serializes the new input vector generated from the Krylov solver in the
successive iterations.

\subsection{GMRES}
AFMPB uses the Generalized minimal residual (GMRES) from
SPARSKIT \cite{sparskit} to solve the discretized system (\ref{eq:mat}).
SPARSKIT is written in Fortran and does not operate on distributed
memory computers. When the bulk of computation in AFMPB becomes
distributed, the GMRES operation must also become distributed.
For this reason, the new version of AFMPB also includes an implementation of
restarted GMRES on distributed memory computers. The implementation
follows closely to the one in SPARSKIT, skipping statements on preconditioners
because (\ref{eq:mat}) is a well-conditioned system resulting from the
second kind Fredholm integral equation formulation.

We also note that in previous AFMPB versions, there are separate buffers
to store the Krylov basis and the input to the multipole method. As the input
to the multipole method is simply the last orthonormal basis formed, this
extra storage is eliminated. Additionally, to reduce unnecessary
communications, the Krylov basis are distributed exactly the same way as
the mesh nodes. In other words, each {\tt Node} object has a {\tt gmres}
member that stores the corresponding Krylov basis components.

\section{Software Installation and Job Examples} 
\label{sec:install}

\subsection{Installation}
The new version AFMPB depends on two external libraries: DASHMM and HPX-5.
Version 4.1.0 of HPX-5 can be downloaded from
\url{https://hpx.crest.iu.edu/download}. DASHMM is automatically downloaded by
AFMPB when the application is built.

Users must install HPX-5 on their systems before installing the AFMPB solver.
HPX-5 currently specifies two network interfaces: the ISend/IRecv interface
with the MPI transport, and Put-with-completion (PWC) interface with the Photon
transport. HPX-5 can be built with or without network transports.
Assume that you have unpacked the HPX-5 source into the folder
{\tt /path/to/hpx} and wish to install the library into
{\tt /path/to/install}. Without network support, the following steps
should build HPX-5 and install it.

\begin{enumerate}
\item {\tt cd /path/to/hpx}
\item {\tt ./configure --prefix=/path/to/install}
\item {\tt make}
\item {\tt make install}
\end{enumerate}

To configure HPX-5 with MPI network, simply add {\tt --enable-mpi} to
the configure line. The configuration will search for the appropriate way
to include and link to MPI.

\begin{enumerate}
\item HPX-5 will try and see if {\tt mpi.h} and {\tt libmpi.so} are available
with no additional flags.
\item HPX-5 will test for an {\tt mpi.h} and {\tt -lmpi} in the current
{\tt C\_INCLUDE\_PATH} and {\tt \{LD\}\_LIBRARY\_PATH} respectively.
\item HPX-5 will look for an {\tt ompi} {\tt pkg-config} package.
\end{enumerate}

To configure HPX-5 with the Photon network, one adds {\tt --enable-photon}
to the configure line. HPX-5 does not provide its own distributed
job launcher, so it is necessary to also use either the {\tt --enable-mpi} or
{\tt --enable-pmi} option in order to build support for {\tt mpirun}
or {\tt aprun} bootstrapping. Note that if you are building with the Photon
network, the libraries for the given network interconnect you are targeting
need to be present on the build system. The two supported interconnects
are InfiniBand ({\tt libverbs} and {\tt librdmacm}) and Cray's GEMINI
and ARIES via uGNI ({\tt libugni}). On Cray machines you also need to include
the {\tt PHOTON\_CARGS=``--enable-ugni''} to the configure line so that
Photon builds with uGNI support. Finally, the {\tt --enable-hugetlbfs} option
causes the HPX-5 heap to be mapped with huge pages, which is necessary
for larger heaps on some Cray Gemini machines.

Once the HPX-5 system is installed, assume the AFMPB source has been unpacked
into the folder {\tt /path/to/afmpb}, the package can be easily built using
{\tt cmake} (version 3.4 and above) as follows

\begin{enumerate}
\item {\tt mkdir build; cd build}
\item {\tt cmake /path/to/afmpb}
\item {\tt make}
\end{enumerate}

\noindent The previous commands will automatically download and build the
DASHMM library, so no extra steps are required to satisfy that dependency.

\subsection{Job Examples}
Assume the exectuable is {\tt afmpb} in this section. The solver can be
used simply as
\begin{verbatim}
./afmpb --pqr-file=FILE
\end{verbatim}
which will discretize the molecule using the built-in mesh generation tool
and compute the potentials and solvent energy. 

The other available options to the program are
\begin{description}
\item[--mesh-format=num] Available choices are 0, 1, 2. 0 indicates built-in
mesh, 1 indicates MSMS mesh~\cite{msms}, and 2 indicates TMSmesh~\cite{tmesh2}.
\item[--mesh-file=FILE] Required if the mesh format is not zero.
\item[--mesh-density=num] Specifies mesh density if built-in mesh is selected.
\item[--probe-radius=num] Specifies probe radius if built-in mesh is selected.
\item[--dielectric-interior=num] Specifies the interior dielectric constant.
\item[--dielectric-exterior=num] Specifies the exterior dielectric constant.
\item[--ion-concentration=num] Specifies the ionic concentration.
\item[--temperature=num] Specifies the temperature
\item[--surface-tension=num] Specifies the surface tension coefficient
\item[--pressure=num] Specifies the pressure.
\item[--accuracy=num] Specifies the accuracy of DASHMM, available choices are
3 and 6.
\item[--rel-tolerance=num] Specifies relative tolerance of (\ref{eq:tol}) 
for GMRES solver
\item[--abs-tolerance=num] Specifies absolute tolerance of (\ref{eq:tol}) 
for GMRES solver 
\item[--restart=num] Specifies the maximum dimension of Krylov space before
it restarts.
\item[--max-restart=num] Specifies the maximum number of times GMRES can
restart.
\item[--log-file=FILE] Name of the log file.
\item[--potential-file=FILE] Name of the file containing computed potentials
on each mesh node.
\end{description}
and they are also available by issuing {\tt ./afmpb --help} from the
command line.

Next, we use the 3K1Q molecule to be discussed in the next subsection to
present some sample job scripts. On a cluster using Slurm workload manager,
a job using 512 compute nodes looks like
\begin{verbatim}
#! /bin/bash -l
#SBATCH -p queue
#SBATCH -N 512
#SBATCH -J jobname
#SBATCH -o output
#SBATCH -e error
#SBATCH -t 00:10:00

srun -n 512 -c 48 ./afmpb --pqr-file=3k1q.pqr \
--mesh-format=2 --mesh-file=3k1q.off --log-file=3k1q.out512 \
--accuracy=6 --restart=79 --max-restart=5 --hpx-threads=24
\end{verbatim}
It is worth noting that each compute node of the above cluster is equipped
with two Intel Xeon E5 12-core CPUs with hyperthreading enabled. This means,
Slurm sees 48 cores per compute node and the option {\tt -c 48} is to create
one MPI process per compute node. As there are only 24 physical cores
on each compute node, the option {\tt --hpx-threads=24} means HPX-5 will create
only 24 scheduler threads, one for each physical core. If the above cluster
were using the PBS workload manager, the script will look like
\begin{verbatim}
#! /bin/bash -l
#PBS -l nodes=512:ppn=48
#PBS -l walltime=00:10:00
#PBS -q queue

aprun -n 512 -d 48 ./afmpb ....
\end{verbatim}

\section{Numerical Examples}
\label{sec:result}

In this section, we demonstrate the accuracy and efficiency of
the new AFMPB solver. AFMPB's accuracy is demonstrated by
considering a spherical case with known a analytic solution. AFMPB's efficiency
is demonstrated by applying the solver to two molecule systems:
an aquareovirus virion 3K1Q and a dengue virus 1K4R. The results were
collected from a Cray XC30 cluster at Indiana University.  Each
compute node has two Intel Xeon E5 12-core CPUs and 64 GB of DDR3 RAM.
The HPX-5 runtime was configured with the Photon network.

The error tolerance for the GMRES solver is chosen to be
\begin{equation}
\label{eq:tol}
\epsilon_{rel} \|b \|_2 + \epsilon_{abs},
\end{equation}
where $b$ represents the right-hand side of (\ref{eq:mat}),
and the relative and absolute tolerance
 $\epsilon_{rel}$ and $\epsilon_{abs}$ depend on DASHMM's accuracy, and
the molecule system. When DASHMM computes with 6-digit accuracy, the lower
bound for $\epsilon_{rel}$ is $10^{-6}$. When DASHMM computes with
3-digit accuracy, the lower bound for $\epsilon_{rel}$ is  $10^{-3}$.

The sphere example used a sphere shaped molecule of 50\AA \, radius and a single
atom carrying 50 elementary charge. The ionic concentraion is set to 0 mM. Under
this assumption, the analytical solution of the polar energy is
$-4046\, kcal/mol$. For the numerical solution, the surface is discretized
into $1,310,720$ triangle elements and $655, 362$ nodes.
The results computed from the AFMPB solver are
$-4065.42\, kcal/mol$ when DASHMM operates with 3-digit accuracy and
$-4057.83\, kcal/mol$ when DASHMM oeprates with 6-digit accuracy. The relative
differences are 0.47\% and 0.29\%, respectively.

\begin{figure}
\centering
\begin{subfigure}[b]{0.45\textwidth}
  \includegraphics[width=\textwidth]{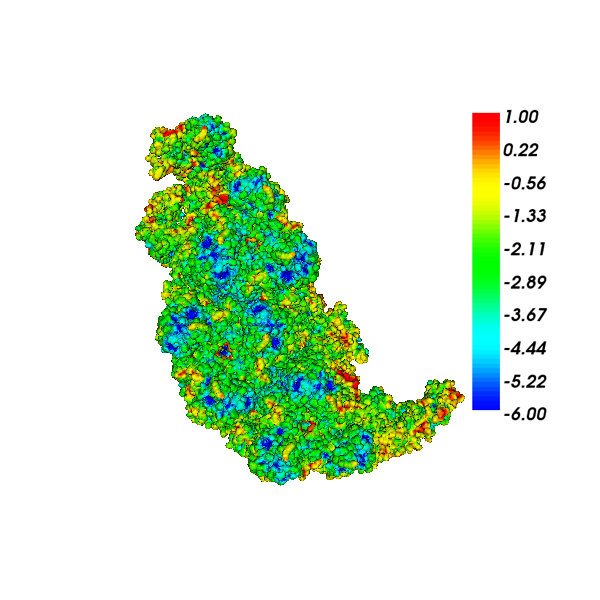}
  \caption{3K1Q}
  \label{fig:3k1q}
\end{subfigure}
~
\begin{subfigure}[b]{0.45\textwidth}
  \includegraphics[width=\textwidth]{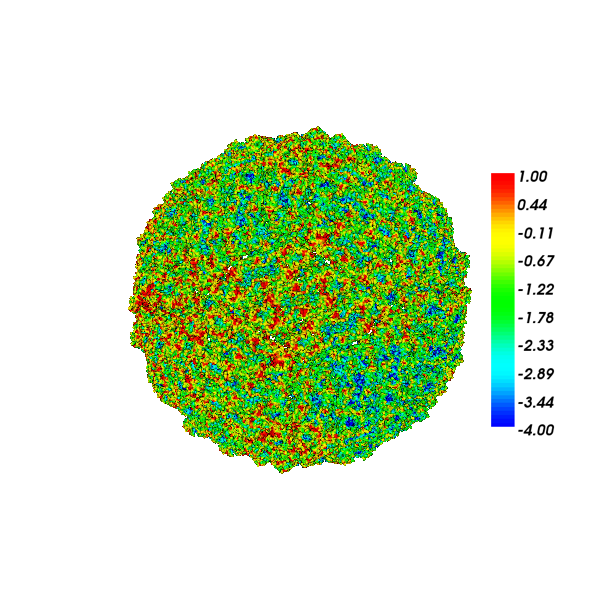}
  \caption{1K4R}
  \label{fig:1k4r}
\end{subfigure}
\caption{Visualization of the surface potential of the aquareovirus
virion (left) and dengue virus (right) systems using VCMM~\cite{vcmm}.
The surface mesh is generated using TMSmesh and the color bar for the surface
potentials is in the unit of $kcal/mol.e$.}
\label{fig:visual}
\end{figure}

The surface mesh used for the 3K1Q and 1K4R examples
were generated using TMSMesh~\cite{tmesh2}.
The smaller molecule, 3K1Q, consists of $203,111$ atoms and its
mesh consists of $7,788,246$ triangle elements and $3,888,281$ nodes.
We were able to test the AFMPB solver on this molecule with both three
and six digit accuracy requirements on DASHMM. At 3-digits of accuracy,
the polar energy is $-8.81762 \times 10^4\, kcal/mol$. At 6-digits of accuracy,
the polar energy is $-8.81808 \times 10^4\, kcal/mol$. The relative difference
between the two accuracy requirement is 0.005\%, indicating that the lower
accuracy input on DASHMM is also acceptable for energy calculations.
 The larger molecule, 1K4R,
consists of $1,082,160$ atoms and its mesh
consists of $19,502,784$ triangle elements and $9,758,426$ nodes. For
this molecule, we required three digits of accuracy and  the polar energy
is $-3.999067\times10^5 \, kcal/mol$.
The output from the AFMPB package can be visualized by the package
VCMM~\cite{vcmm}. Figure~\ref{fig:visual} shows the visualization of
the surface potentials of the two molecule systems. The potential results
for the 3K1Q molecule on the left were computed with 6-digit accuracy on
DASHMM.

To evaluate the strong scaling performance of the new AFMPB solver, we
measure the execution time of the DASHMM evaluation phase and the
GMRES phase. The DASHMM phase accumulates the total time spent on
matrix-vector multiplication and the GMRES phase accumulates the time
spent on inner production computation.  Because the GMRES
implementation is based on the modified Gram-Schmidt procedure with
reorthogonalization, each inner product is a global
barrier. Therefore, we also count the number of inner products
performed in each solve.  For each accuracy requirement, we started
with the smallest number of compute nodes that could complete the
computation, and used up to 512 compute nodes.  GMRES was set to
restart after 80 iterations and was allowed to restart five times.  In
all test cases, the solver converged well before that limit.  For
3K1Q, GMRES took 11 iterations and 77 inner products to converge when
DASHMM gave 3-digit accuracy, and 133 iterations and $4,803$ inner
products to converge when DASHMM gave 6-digit accuracy. For 1K4R,
GMRES took 10 iterations and 66 inner product to converge.

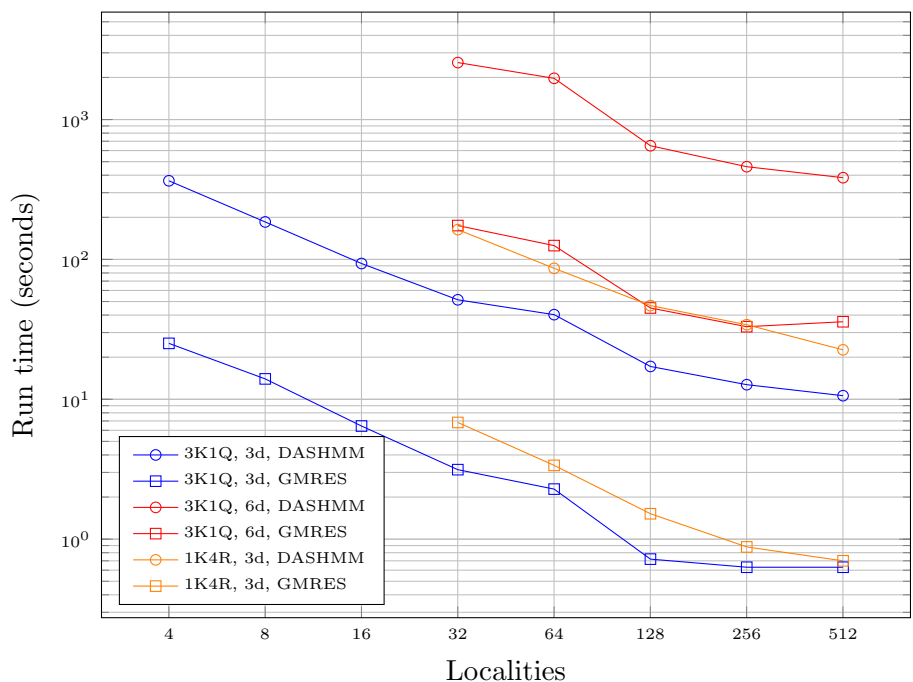
\begin{figure}
\centering
\begin{tikzpicture}
  \begin{loglogaxis}[     
      ylabel = {\small Run time (seconds)},  
      xlabel = {\small Localities}, 
      width = 0.9\textwidth, 
      height = 0.5\textheight, 
      log basis x = 2, 
      xticklabel=\pgfmathparse{2^\tick}\pgfmathprintnumber{\pgfmathresult},
      tick label style={font=\tiny}, 
      grid = both, 
      legend cell align=left, 
      legend entries = {3K1Q, 3d, DASHMM \\
        3K1Q, 3d, GMRES\\
        3K1Q, 6d, DASHMM \\
        3K1Q, 6d, GMRES\\
        1K4R, 3d, DASHMM\\
        1K4R, 3d, GMRES\\
      }, 
      legend columns= 1, 
      legend style={font={\tiny}, at={(0.35,0.3)}, inner xsep=5pt}, 
    ]
    \addplot [blue, mark=o] coordinates {
      (4, 364.97)
      (8, 185.33)
      (16, 93.48)
      (32, 51.45)
      (64, 40.29)
      (128,17.17)
      (256, 12.72)
      (512, 10.61)
    };
    \addplot [blue, mark=square] coordinates {
      (4, 25.11)
      (8, 13.99)
      (16, 6.44)
      (32, 3.14)
      (64, 2.28)
      (128, 0.72)
      (256, 0.63)
      (512, 0.63)
    }; 
    \addplot [red, mark=o] coordinates {
      (32, 2555.13)
      (64, 1971.84)
      (128, 649.98)
      (256, 460.18)
      (512, 383.99)
    }; 
    \addplot [red, mark=square] coordinates {
      (32, 174.66)
      (64, 125.55)
      (128, 44.96)
      (256, 33.04)
      (512, 35.85)
    }; 
    \addplot [orange, mark=o] coordinates {
      (32, 163.37)
      (64, 86.40)
      (128, 46.76)
      (256, 34.13)
      (512, 22.57)
    };
    \addplot [orange, mark=square] coordinates {
      (32, 6.83)
      (64, 3.37)
      (128, 1.52)
      (256, 0.88)
      (512, 0.7)
    }; 
  \end{loglogaxis}
\end{tikzpicture} 
\caption{Execution times for DASHMM (matrix-vector multiplication)
and GMRES (inner-product) for the three test cases on a Cray XC30 cluster.
Each compute node (locality) consists of 24 cores and 64 GB of RAM.
 Molecule 3K1Q was tested with two different accuracy requirements for DASHMM.
 At 3-digit accuracy,
GMRES took 11 iterations and 77 inner products to converge. At 6-digit
accuracy, GMRES was restarted after 80 iterations and took 133 iterations and
4803 inner products to converge.  Molecule 1K4R was tested with
three digit accuracy requirement for DASHMM. The GMRES took 10 iterations
and 66 inner products to converge.}
\label{fig:time}
\end{figure}

Figure~\ref{fig:time} shows the strong-scaling behavior for both the DASHMM
and GMRES phases. Notice that the results for the GMRES phase depend on
how DASHMM distributes the data. For instance, DASHMM didn't do an optimal
job distributing the data at 64 ranks for 3K1Q because the molecular
surface is highly irregular (Figure~\ref{fig:visual}).
  However, DASHMM distributed the data very well at
128 ranks as the strong scaling efficiency for GMRES is almost 100\%.
Nonetheless, each inner product is
a global reduction and as the number of localities increases the computation
will become insufficient to hide the network latency. This explains
why all the GMRES curves flatten out at 512 localities.

\begin{table}
\centering
\begin{tabular}{|c|c|c|c|}
\hline
Localities & 3K1Q, 3-digit & 3K1Q, 6-digit & 1K4R, 3-digit \\
\hline
8 & 98\% & N/A & N/A \\
\hline
16 & 97\% & N/A & N/A \\
\hline
32 & 89\% & N/A & N/A \\
\hline
64 & 57\% & 65\% & 95\% \\
\hline
128 & 66\% & 98\% & 87\% \\
\hline
256 & 45\% & 69\% & 60\% \\
\hline
512 & 27\% & 42\% & 45\% \\
\hline
\end{tabular}
\caption{Strong scaling efficiency of the DASHMM part for 3K1Q and 1K4R
test cases.}
\label{tab:scale}
\end{table}

Table~\ref{tab:scale} summarizes the strong-scaling efficiency for the
DASHMM phase of the AFMPB solver. Note that in all test cases, GMRES
takes approximately 5\% of the total execution and therefore the overall
scaling is determined by that of DASHMM. For 3K1Q at lower accuracy requirement,
the scaling efficiency is very good up to 32 localities and then decays
rapidly. In fact, an input size around $4$ million points is very small and
a single DASHMM evaluation could even fit in a single locality. Running
it over distributed memory primarily adds memory capacity for the GMRES
solver to work.  For higher accuracy requirement
or large molecule, the scaling efficiency can stay above 60\% up to 256
localities, which is consistent with the ones reported in \cite{debuhr2017b}.
However, the efficiency at 512 localities are much worse.

The inferior scaling performance at 512 localities is caused by the fact that
AFMPB does not distinguish types 1 and 3 lists in the adaptive FMM (see
Section~\ref{subsec:ext}). In the adaptive FMM, if a source box
$B_s$ is in list-3 for target box $B_t$, their interaction is handled by
the {\tt M\_to\_T} operator. If $B_s$ and $B_t$ are on different localities,
one needs to communicate the multipole expansion of $B_s$ and the
{\tt M\_to\_T} operator has enough floating point operations (such as
spherical harmonic evaluations) to offset the communication cost.
In AFMPB, this interaction is handled by the {\tt S\_to\_T} operator and
$B_s$ needs to send ``particles'' (component of the Krylov basis) to $B_t$.
First, $B_s$ can be a nonleaf box which contains many particles, making the
message much larger. Second, once the $\{A, B, C, D\}$ coefficients are
computed, for each particle information received, $B_t$ simply does four
multiplications, which is not sufficient to offset the communication cost.
The DASHMM team is currently working on extending the library to heterogeneous
architectures. When complete, offloading the near-field interaction
should be able to improve the scaling performance.

Finally, we point out that one often chooses restarted GMRES due to the
memory limitation on storing the Krylov basis and it often takes more iterations
to converge when GMRES restarts. When there are more resources available,
one can afford not to restart GMRES and this could lead to shorter
time-to-completion. For the 3K1Q example at 6-digit accuracy, we set GMRES
to restart at 140 iterations and GMRES actually converged at iteration 89
and the execution time is 30\% faster.
The polar energy is $-8.83353\times10^4\, kcal/mol$.
Compared with the one obtained
when GMRES restarted, the relative difference is less than 0.17\%.


\section{Conclusion}
\label{sec:conclusion}

We have presented the DASHMM accelerated AFMPB solver for computing
electrostatic properties and solvation energies of biomolecular
systems. DASHMM leverages the global address space of the HPX-5 runtime
system to provide a unified evaluation of the multipole methods on
both shared and distributed memory computers. This enables the new version
of AFMPB to operate on distributed memory computers while at the same time
maintaining  backward compatibility on shared memory computers. With the added
processing units and memory capacity from distributed memory computers,
larger molecules or situations requiring higher accuracy can be
handled. We have tested the new solver on the dengue
virus system reported in the previous release and was able to reduce the
solving time from 10 hours down to about 30 seconds using $12,288$ cores.

The adoption of DASHMM as the FMM solver for AFMPB also revealed several issues
requiring further study. A better treatment of the near-field interaction is
needed to recover the scaling demonstrated by DASHMM in other contexts.
One possible approach would be the use of accelerators for near field
interactions. The current implementation of the GMRES solver follows closely
the one in SPARSKIT. The modified Gram-Schmidt procedure involves many global
synchronizations. Given the current breakdown of execution time devoted to
FMM and GMRES, it should be possible to hide this synchronization overhead
withing the matrix-vector multiplication using strategies mentioned in
\cite{Ghysels2013} and references therein.

\section*{Acknowledgments}
Author Lu acknowledges the support of Science Challenge Project
(No. TZ2016003) and China NSF (NSFC 21573274) in China. Authors Zhang,
DeBuhr, and Sterling were supported in part by National Science
Foundation grant number ACI-1440396. Authors Niedzielski and Mayolo
gratefully acknowledge the support of the National Science
Foundation's REU program. This research was supported in part by Lilly
Endowment, Inc., through its support for the Indiana University
Pervasive Technology Institute.

\bibliographystyle{elsarticle-num}
\bibliography{afmpb}

\end{document}